\documentclass{elsarticle}

\usepackage{lineno,hyperref}
\modulolinenumbers[5]

\journal{arXiv}

\begin{document}

\begin{frontmatter}

\title{Anti-Tamper Protection for Internet of Things System Using Hyperledger Fabric Blockchain Technology}

\author{Adnan Iftekhar}
\address{Key Laboratory of Aerospace Information Security and Trusted Computing, Ministry of Education, Wuhan University, Wuhan 430000, China(e-mail: adnan@whu.edu.cn)}

\author{Xiaohui Cui}
\address{Key Laboratory of Aerospace Information Security and Trusted Computing, Ministry of Education, Wuhan University, Wuhan 430000, China(e-mail: xcui@whu.edu.cn)}

\cortext[mycorrespondingauthor]{Corresponding author}
\ead{xcui@whu.edu.cn}

\begin{abstract}
Automated and industrial Internet of Things (IoT) devices are increasing daily. As the number of IoT devices grows, the volume of data generated by them will also grow. Managing these rapidly expanding IoT devices and enormous data efficiently to be available to all authorized users without compromising its integrity will become essential in the near future. On the other side, many information security incidents have been recorded, increasing the requirement for countermeasures. While safeguards against hostile third parties have been commonplace until now, operators and parties have seen an increase in demand for data falsification detection and blocking. Blockchain technology is well-known for its privacy, immutability, and decentralized nature. Single-board computers are becoming more powerful while also becoming more affordable as IoT platforms. These single-board computers are gaining traction in the automation industry. This study focuses on a paradigm of IoT-Blockchain integration where the blockchain node runs autonomously on the IoT platform itself. It enables the system to conduct machine-to-machine transactions without the intervention of a person and to exert direct access control over IoT devices. This paper assumed that the readers are familiar with Hyperledger Fabric basic operations and focus on the practical approach of integration. A basic introduction is provided for the newbie on the blockchain.
\end{abstract}


\end{frontmatter}


\section{Introduction}
\label{sec: introduction}

The German Government initiated the fourth industrial revolution (Industry 4.0) in 2013 \cite{wang2021blockchain}. It seeks to create an intelligent production system utilizing economically, ecologically, and socially sustainable cyber-physical systems (CPS) and human equipment interfaces  \cite{bahrin2016industry}. Through big data analytics, the Internet of Things, robotic systems, and artificial intelligence, Industry4.0 aims to increase automation, interoperability, actionable insights, and information transparency \cite{haleem2019additive}. IoT has gained significant traction in the automation sector. The Internet of Things or IoT is a wide term that refers to a variety of intelligent sensors and microcontrollers that gather data from their surroundings. These sensors serve as the foundation for contemporary technologies such as smart homes, smart cities, smart grids, intelligent health systems, and wearable technology \cite{nivzetic2020internet}. IoT applications provide a smooth cyber world connection with the real world. Despite its vast potential, various applications, rapid growth, and future aspirations, IoT currently confronts challenges. Among the significant impediments are data privacy, security, and centralization \cite{mohanta2020survey}. Existing Internet of Things systems are often built on a centralized architecture, with cloud-based servers and edge servers serving as data analysis and information processing nodes, respectively. Despite of efficient administration and strong computing capabilities, the security and privacy concerns connected with such systems are becoming more prevalent.

Collaboration in centralized systems requires multiple parties to either trust one another or pay a third party to facilitate the collaboration, preventing the interoperability of different Internet of Things apps and services \cite{tyagi2020review}. Alternatively, if achieved, decentralization would have a number of advantages to centralized infrastructure, including lower costs. The most important result of decentralization in the Internet of Things is the development of distributed consensus among IoT devices \cite{bodkhe2020survey}. If properly controlled, it can significantly improve the IoT system security and offer customers more privacy through adequate data protection measures. Decentralized Internet of Things solutions, which do not need the participation of a trusted central authority, are capable of managing a large number of transactions and scaling to many peers in order to reach agreement \cite{e23081054}. Notably, the creation of the blockchain has enabled large-scale applications to overcome distributed consensus limitations in a decentralized environment \cite{panda2019study}. The IoT industry's security and privacy problems can be addressed through blockchain technology \cite{wang2020blockchain}. At its core, it is used to store a variety of different cryptocurrency transactions. However, it is not restricted to cryptocurrencies only. Additionally, it may be used to store various other types of data, including assets, IoT data, and multimedia data \cite{bhowmik2017multimedia}. 

Academia and industry have both shown a significant deal of interest in emerging blockchain technology. Blockchain technology is a distributed, unchangeable ledger that simplifies the method of documenting transactions and keeping track of assets over a peer-to-peer network of computers \cite{nofer2017blockchain}. It can record tangible and intangible assets like intellectual property, patents, copyrights, and many more. Almost everything of value can be monitored and sold on a blockchain network, which mitigates risk and lowers costs for all parties by eliminating the intermediaries. In this article, we use Hyperledger Fabric Blockchain technology to protect an IoT network.

The main contributions of this paper are:

\begin{itemize}
  \item We examine the typical IoT architecture and assess the security and privacy concerns associated with IoT systems.
  \item We formulate how blockchain may be used in conjunction with IoT and offer a framework for integrating blockchain and IoT.
  \item The suggested system uses a blockchain-based access control system to maintain a separation between users and gadgets.
\end{itemize}

The rest of the article is organized as follows. In Section \ref{sec:preliminary} we discussed the overview of the blockchain and related studies on blockchain-IoT integration. We discussed some potential blockchain-IoT configurations used in the industry. In Section III, we proposed our IoT-Blockchain integration model. We discussed the Prototype of our proposed model in Section IV, while section V describes its working methodology. Section VI discussed the Results and advantages of our proposed configuration, followed by the Conclusion and future work.

\section{Preliminary}
\label{sec:preliminary}

Throughout this section, we covered some fundamental blockchain concepts. We provided a brief description of Hyperledger Fabric blockchain technology and the work in integrating IoT and blockchain.

\subsection{Blockchain Overview}

Distributed Ledger Technology (DLT) is a term that refers to a decentralized database that is managed by a large number of individuals \cite{sunyaev2020distributed}. The term "blockchain" refers to a form of distributed ledger technology in which transactions are recorded using an immutable cryptographic signature known as a hash \cite{kuznetsov2019statistical}. It is simple to understand the concept of a blockchain as a distributed digital record of transactions that is duplicated and disseminated across the encrypted blockchain's network of peer-to-peer computer systems \cite{androulaki2018hyperledger}. It is almost difficult to modify, hack, or cheat the blockchain system in any significant way. Each block in the chain is made up of a number of different transactions. On the blockchain, every time a new transaction takes place, a record of that transaction is recorded in the ledger of each participant.

\begin{figure}[h]
\centerline{\includegraphics[scale=0.60]{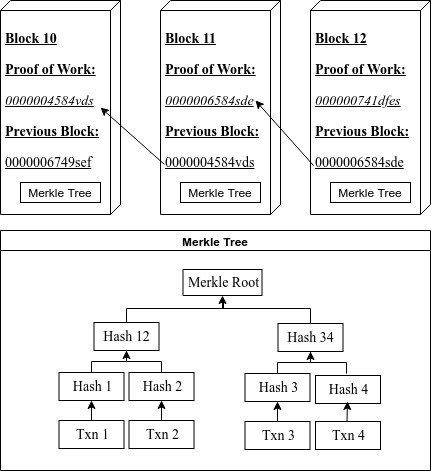}}
\caption{Blocks of a blockchain system}
\label{fig:blocks}
\end{figure}

When the block of a blockchain is formed, the cryptographic hash is generated using a nonce. Unless the block is mined, its data is deemed signed and will remain permanently associated with the nonce and hash. Every block in a blockchain has its nonce and hash, but it also references the preceding block's hash [Figure \ref{fig:blocks}].

This means that if a single block in a chain is changed, it is instantly obvious that blockchain has been modified. In order to compromise a blockchain system, hackers would have to alter every block across all distributed versions and regenerate the hash of that block and all the following blocks. Blockchain networks are classified into three types: public, private, and consortium blockchains \cite{iftekhar2020application}. Each of these platforms has several advantages, disadvantages, and optimal usage. A brief description of blockchain types is given in Table \ref{tab:classificationofblockchain}.

\begin{table}[h]
\centering
\caption{{Classification of Blockchains} \cite{iftekhar2021hyperledger}}
\label{tab:classificationofblockchain}
\scalebox{0.62}{
\begin{tabular}{llll} 
\hline
& \textbf{Public Blockchain}                                                          
& \textbf{Private Blockchain}                                                             
& \textbf{Permissioned Blockchain}
\\ 
\hline
\textbf{Read Access}       
& \begin{tabular}[c]{@{}l@{}}No permission required \\from any authority\end{tabular} 
& \begin{tabular}[c]{@{}l@{}}Read Access is private within \\organization participants\end{tabular}  
& \begin{tabular}[c]{@{}l@{}}Public/Participants are permissible \\under some legal contract\end{tabular}  
\\
 &                                                                                     
 &                                                                                        
 &                                                                                      
 \\
\textbf{Write Access}      
& \begin{tabular}[c]{@{}l@{}}No permission required \\from any authority\end{tabular} 
& \begin{tabular}[c]{@{}l@{}}Write Access is private within \\organization participants\end{tabular} 
& \begin{tabular}[c]{@{}l@{}}Participants are permissible under \\some legal contract\end{tabular}         
\\
&                                                                                     
& 
&      
\\
\textbf{Consensus Process} & \begin{tabular}[c]{@{}l@{}}Anyone can join \\consensus process\end{tabular}         
& \begin{tabular}[c]{@{}l@{}}Pre-selected nodes within \\organization\end{tabular}       
& \begin{tabular}[c]{@{}l@{}}Pre-selected nodes within \\consortium\end{tabular}        
\\
\hline
\end{tabular}}
\end{table}


\subsection{Hyperledger Fabric Blockchain}

Fabric is a platform for building business Blockchain applications \cite{androulaki2018hyperledger}. It is open-source and developed by IBM under the Linux Foundation. It is a consortium chain; therefore, it is excellent for businesses that do not want their transactions public and needs faster transactions. Unlike Bitcoin, Fabric is a permissioned Blockchain that follows a set of protocols \cite{ABlockch60:online}.

\subsubsection{Peers} 

Peers are the basic building blocks of any distributed paradigm \cite{Peers:online}. Although peers are simply nodes, they are all equal on public blockchain networks such as Bitcoin or Ehtereum, but peers are not equal consortium Blockchains. Each organization in the blockchain network must set up peer nodes. These peers are all connected to the blockchain network through specific permissions and signatures. Peers are essential to the network since they host the ledger and chaincode. A peer may host a ledger instance without hosting any chaincodes that access it. 

Each organization's peer has a unique set of behaviors and tasks. Here are the kinds of peers.

\subsubsection{Anchor Peers}
The peers of one organization are known to other organisations, so that they may interact with each other by means of a protocol via Gossip termed the peers of Anchor. These peers can't be found by other organizations termed Non-Anchor peers.

\subsubsection{Endorsement Peers} 

The endorsement policy is determined by who does the endorsing and what constitutes a valid transaction. The endorsement policy includes a list of peers who will be involved in signing transaction proposals \cite{Endorsem87:online}. Whenever a network is established at that moment, we have to select which peers should endorse the transaction or verify the transaction. 

\subsubsection{Committing Peers} 

Committing peers are peers that do not require the ChainCode to be deployed but are responsible for maintaining the entire ledger of the records.

\subsubsection{Orderer Service} 

Ordering Peers collect approved transactions, bundle them into blocks, and distribute them to other peers for validation \cite{orderer:online}. While ordering, nodes maintain a record of all transactions in their ledger, including valid and invalid transactions, endorsing peers, and committing peers to maintain a record of just valid transactions.

\subsubsection{Chaincode} 

Chaincodes are the Fabric Smart Contracts that run the network's business logic. Chaincode is the asset's storing software. Assets are a key-value combination that allows the network to trade practically anything with monetary worth. Chaincode is a Docker container application and can be developed in Go or Java programming language \cite{Chaincod87:online}.

\subsubsection{Channel}
A Hyperledger Fabric channel is a private "subnet" of communication used by two or more particular network participants to perform private and confidential transactions \cite{Channels47:online}. Members (organizations), anchor peers per member, the shared ledger, chaincode application(s), and the ordering service node all contribute to the definition of a channel (s). Each transaction on the network takes place via a channel, which requires each participant to be authenticated and allowed to transact on. Each peer that enters a channel is assigned a unique identity by a membership services provider (MSP), which authenticates each peer with other peers and services in the channel.

\subsubsection{Membership Service Provider (MSP)} 

All the components of Hyperleger Fabric require valid credentials to join the network. MSPs are a built-in component that gives all the network actors a credential in an X509 certificate. Clients use these credentials to authenticate transactions, while peers use them to authenticate results.

Technology selection is a long-term investment for a business. The great degree of flexibility allowed by open source code, modular components, and standard compliance is one of the main advantages of an open source business application. This allows a business to quickly change technology in order to achieve true usability. For instance, if an organization currently uses kafka clusters as their messaging protocol, we may use it as the consensus mechanism in the Hyperleger Fabric blockchain network instead of the RAFT orderering service. Additionally, we may use existing consensus algorithms or build our own lightweight consensus algorithm for IoT-platforms in order to overcome processing power limitations. Enterprise and corporate networks are not obliged to employ a single vendor, but may instead choose from the most creative and active communities, taking advantage of the blockchain technology sector's fast pace of innovation.

\subsection{Blockchain and IoT Integration}

Although blockchain ensures the immutability of data recorded on the distributed ledger, it does not have the power to verify the validity of data provided by supply chain actors \cite{malik2019trustchain}. To assure data provenance and integrity, the integration of IoT devices and blockchain technology is being advocated\cite{malik2019trustchain,liu2017blockchain}. The danger of malicious data tampering is reduced when data is saved by IoT devices rather than people on the blockchain system. It increases the data's authenticity and trustworthiness. In this sense, combining blockchain with IoT is viewed as a viable way to handle data-related concerns of trust, security, privacy, and integrity \cite{cao2019internet}.

This article offers a permissioned blockchain architecture and protocol suite for a local IoT network. The design is built on a sealed Sequencer and a Fog Server that runs Guy Fawkes protocols (post-quantum). Along with any user data and validity proofs, the blockchain's blocks are kept in networked Content Addressable Storage \cite{shafarenko2021pls}.

Authors provided a blockchain-based system for differentially private data publishing in this study. Two techniques were proposed for disseminating histograms and anonymised data, respectively. By encrypting the collector's and requester's interactions on the blockchain, the suggested protocols can prohibit the collector from publishing faked data and the requester from renouncing the payment \cite{xu2020blockchain}.

To connect huge power Internet of Things devices at the edge of a 5G network, the authors merge blockchain technology with 5G Mobile Edge Computing \cite{wang2021blockchain}. The author explored the method for integrating blockchain and IIoT from an industrial perspective and developed a blockchain-enabled IIoT architecture for establishing trust between IIoT components. Additionally, the authors built a smart contract system for storing and processing data about the interaction of IIoT model components \cite{zhao2019blockchain}.

The authors presented a system for secure and decentralized IoT data exchange that integrates IoT platforms and blockchain, with the Hyperledger Fabric architecture serving as the blockchain back-end. The blockchain is used to store and make access control decisions. The framework implemented cryptography-enforced access control through the use of identity-based encryption \cite{truong2019towards}.

The authors presented a blockchain-based cross-domain authentication procedure for use in an IoT setting. The PBFT algorithm is used in this process. The authentication process is carried out via a smart contract. To authenticate and share authentication data securely, an encrypted key sharing technique is utilized \cite{li2020research}.

The authors presented a decentralized access control system for the Internet of Things based on Blockchain technology (IoT). Permission assignment and access control for cross-domain users/IoT devices are handled via hierarchical smart contracts. To search for and obtain user/IoT device platform hashes, the technique employed a Proof-of-Authenticity/Integrity (PoAI) approach \cite{ali2020xdbauth}.

This study addressed an IoT device access control system built on Hyperledger Fabric. By implementing rules and programmatic access management in the chaincode, the system separates people and devices. A Raspberry Pi 4B node was used to show the fabric policies, chaincode installation, chaincode invocation, and solution benchmarking. According to the research, Hyperledger Fabric blockchain technology comes pre-loaded with IoT device management features \cite{iftekhar2021hyperledger}.

Using a blockchain-based architecture, the authors suggested a decentralized data integrity verification system for Internet of Things data stored in a semi-trusted cloud \cite{liu2017blockchain}. To guarantee data storage, monitoring, and verification are all reliable, the authors proposed an integrated approach that makes use of blockchain technology to do so. The suggested virtual machine agent model is implemented in the cloud utilizing mobile agent technology \cite{wei2020blockchain}.

\subsection{Discussion}

\begin{figure}[h]
\centerline{\includegraphics[scale=0.23]{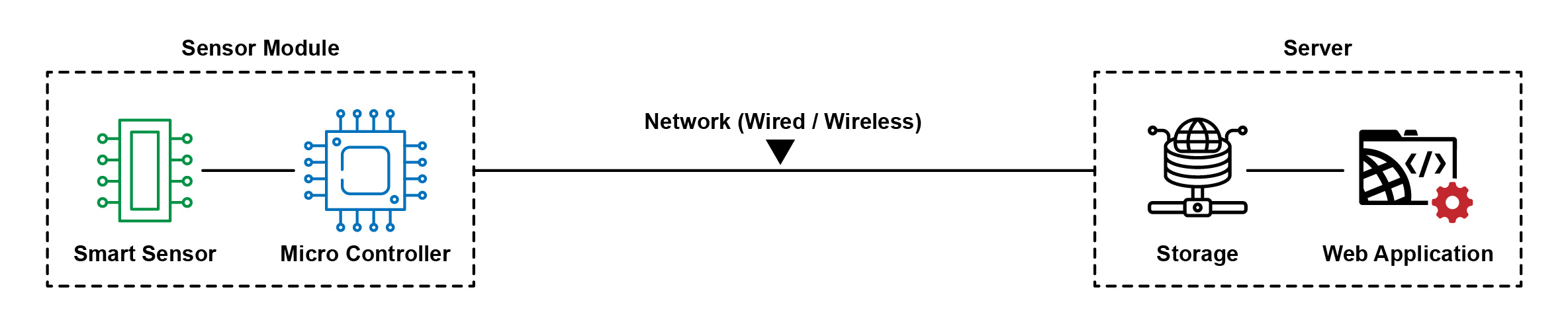}}
\caption{Simple IoT Configuration Package}
\label{fig:iot-config-a}
\end{figure}

Based on their intended application, several types of IoT modules are developed. As shown in Figure \ref{fig:iot-config-a}, the simplest possible connection between sensors and the application server is established. In this kind of setup, the microcontroller to which the sensor is connected can directly communicate with a server through a network. The data from the sensor is sent to the server via the controller, and the server that receives it records it in the blockchain, after which the relevant application consumes the data. In another configuration, like ESP8266 or ESP12F controllers, the microcontroller may expose the web API to the outside world. It is the case where an application gets data from a controller and stores it on a blockchain, where the relevant application later consumes it.

\begin{figure}[h]
\centerline{\includegraphics[scale=0.22]{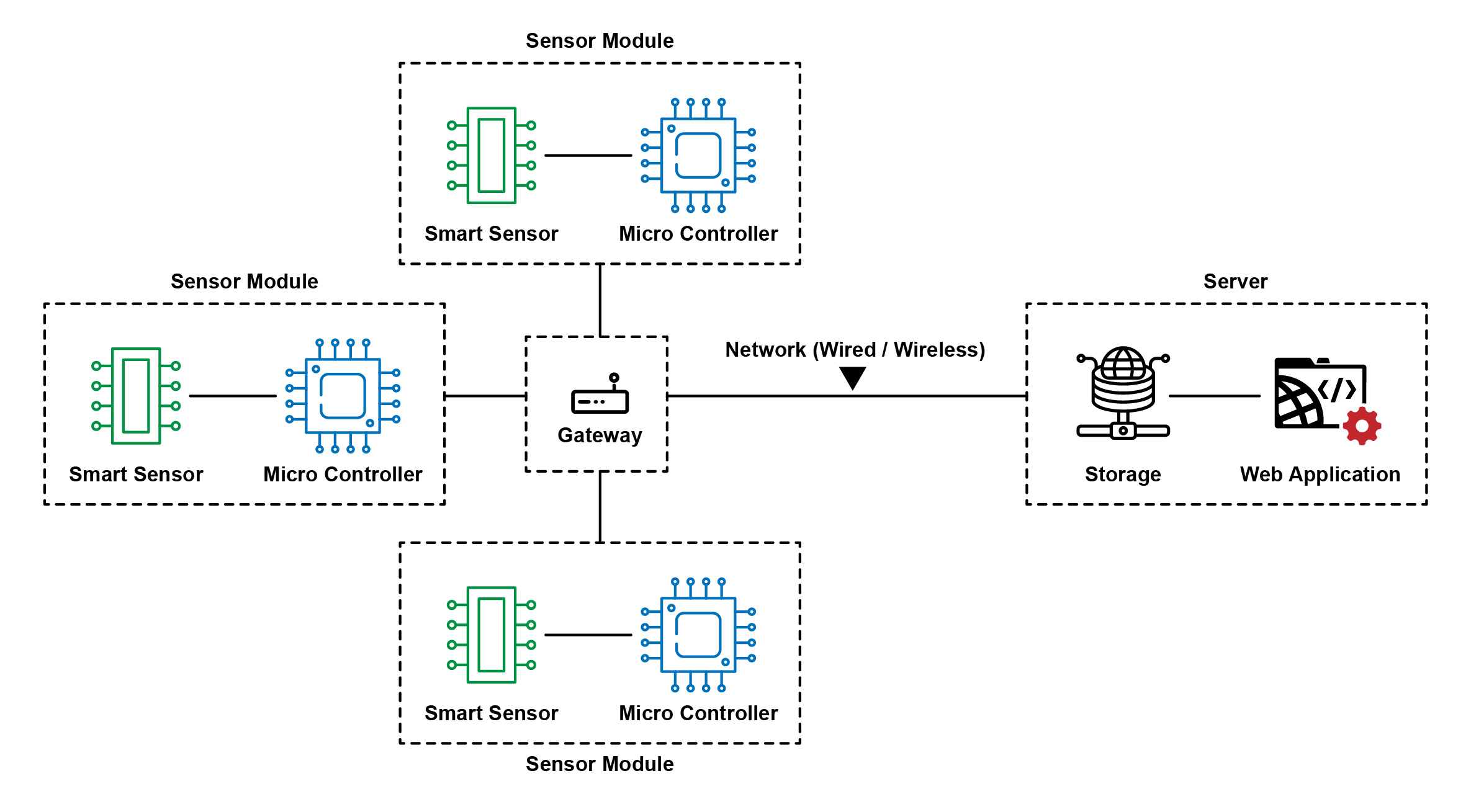}}
\caption{IoT Configuration Package with a Gateway}
\label{fig:iot-config-b}
\end{figure}

When the communication protocol on the sensor module is designed to reduce power consumption and ensure security, a gateway module with a security protocol is included in the communication protocol. This gateway serves as a barrier between the rest of the network and the Internet of Things sensors. In reality, this gateway may serve as an access control protocol for Internet of Things devices occasionally [Figure \ref{fig:iot-config-b}].

\begin{figure}[h]
\centerline{\includegraphics[scale=0.22]{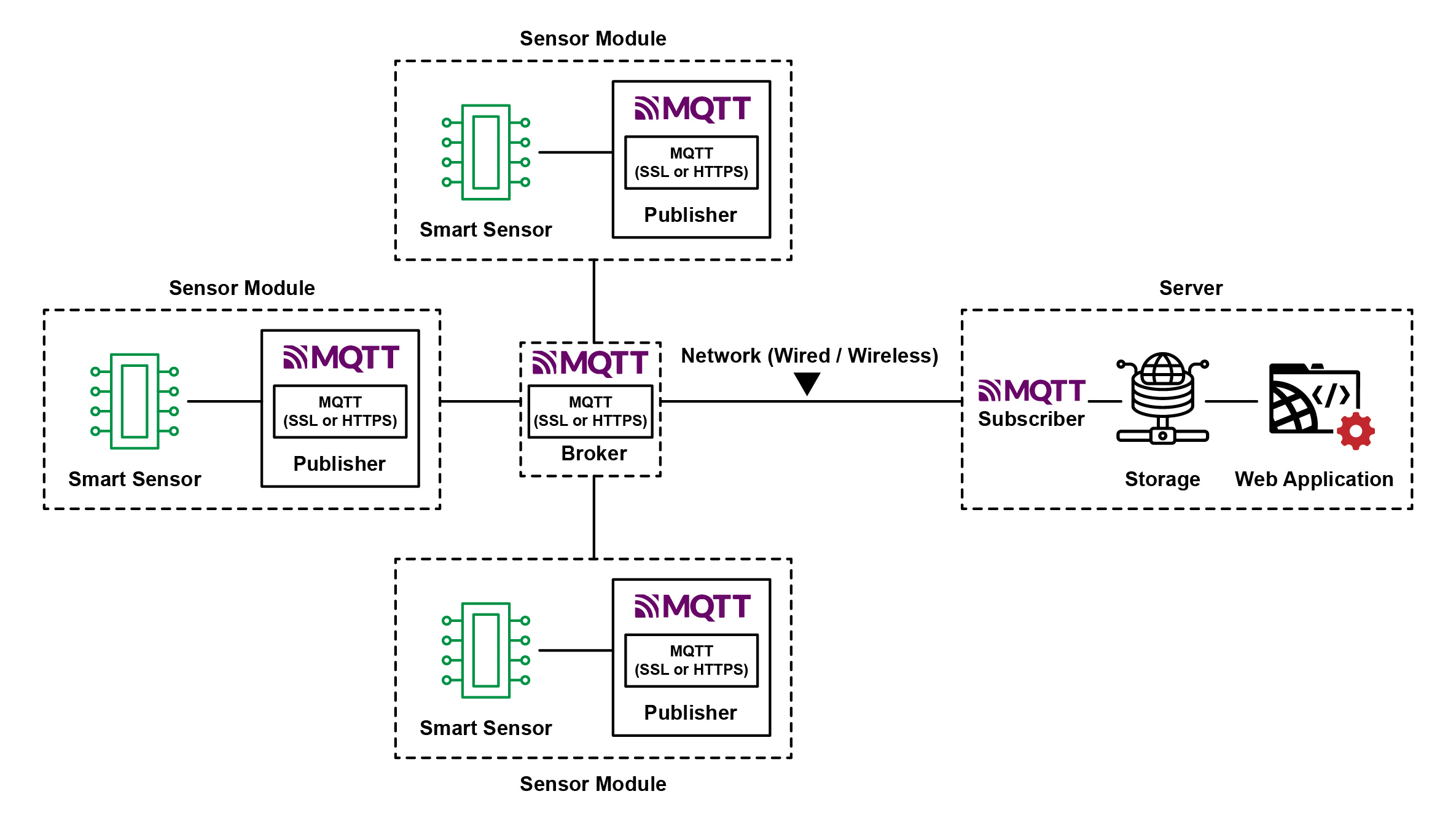}}
\caption{IoT Configuration Package with an MQTT Broker}
\label{fig:iot-config-c}
\end{figure}

The adoption of encryption through the Message Queuing Telemetry Transport (MQTT) protocol is the configuration most often used to protect the online IoT network from being compromised by hackers. The gateway device often performs the function of an MQTT broker. The sensor module sends the data to the gateway over TLS. The respective application consumes that data as a client and puts it on the blockchain network [Figure \ref{fig:iot-config-c}].

The data produced by IoT systems may include trade secrets that are directly linked to the future growth of the organization and that must be kept private from unrelated parties in order for the company to succeed. It is necessary for the data to stay intact. Cloud storage, for example, can be linked with the blockchain system, but it has numerous inherent vulnerabilities. The central server may become vulnerable. Besides more devices with a central server may generate many to one traffic jam and higher latency and a brick point for scalability. Adding devices to a central server may cause traffic jams, increased latency, and scalability issues.

The lack of adequate access control to resources and sensitive information in IoT systems is yet another security concern. A centralized party produces a suitable key based on access rules in traditional authentication and access control management. However, as the IoT system develops, centralized approaches become a bottleneck. The dynamic nature of IoT deployment leads to complicated trust management, compromising system scalability. 

An IoT system gathers data from a range of smart devices and sensors in order to make a complete judgment based on that data and take the actions in an automated systems. However, privacy is readily breached in the complex IoT system during data acquisition, raw data processing, and data sharing. As a result, privacy preservation in Internet of Things systems, both data privacy and entity privacy, is both critical.

\section{IoT-Blockchian Integration Model}

The protection of information throughout the whole Internet of Things ecosystem is a major issue for IoT companies. Customers of the Internet of Things, who are often a group of business partners, need data and insights from IoT devices promptly, at a low cost, and they must be trustable. The use of blockchain technology to store IoT data adds an additional degree of security to help prevent unwanted assaults. DDoS attacks, malicious attacks, and data breaches are all possible with IoT devices. Some IoT devices are constantly connected to power and Wi-Fi, while others are not. No small device can operate a computational and bandwidth-intensive blockchain transaction system. A gateway or similar device, such as Raspberry Pi single board computers, may be required. So these ecosystems must be naturally cooperative.

\begin{figure*}[t]
\centerline{\includegraphics[scale=0.25]{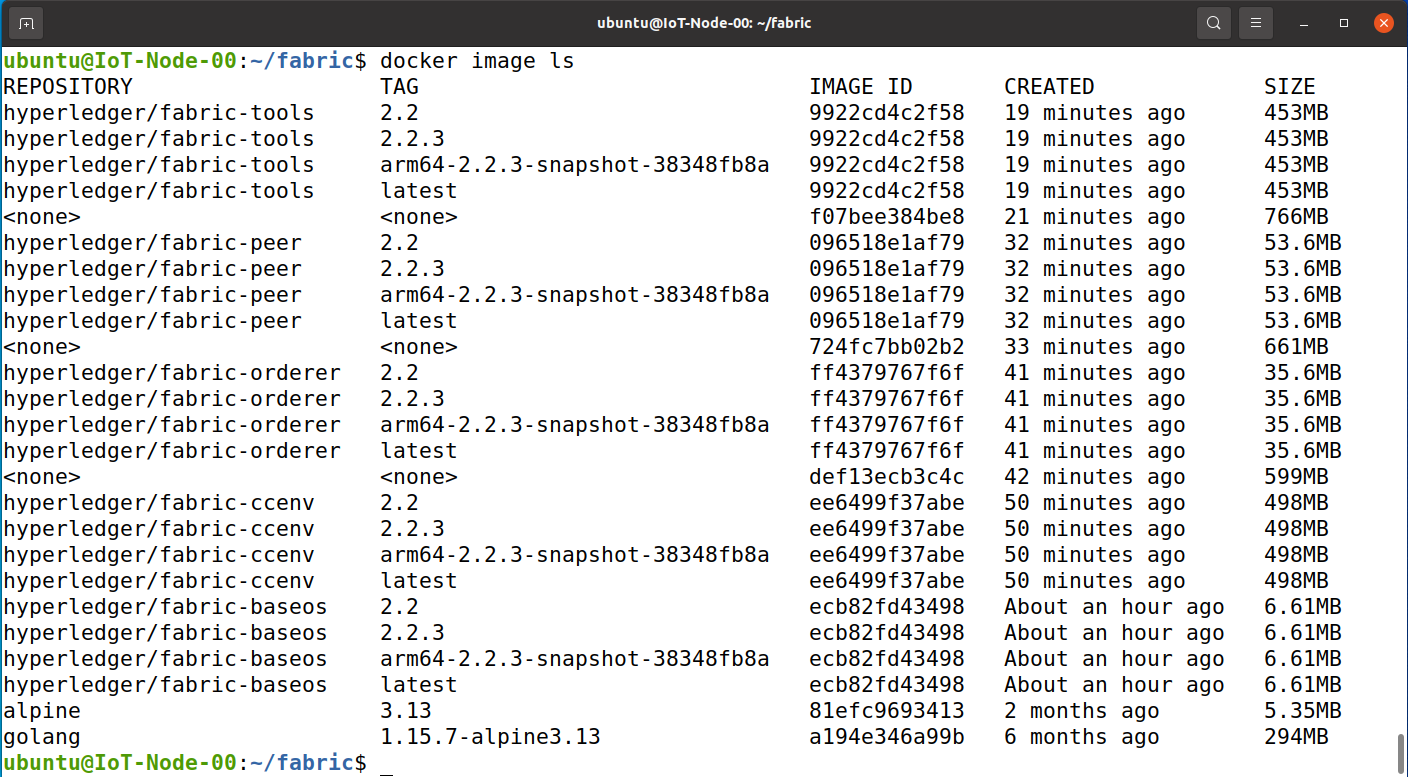}}
\caption{Hyperledger Fabric docker images on Raspberry Pi.}
\label{fig:pi-docker}
\end{figure*}

\begin{figure*}[t]
\centerline{\includegraphics[scale=0.24]{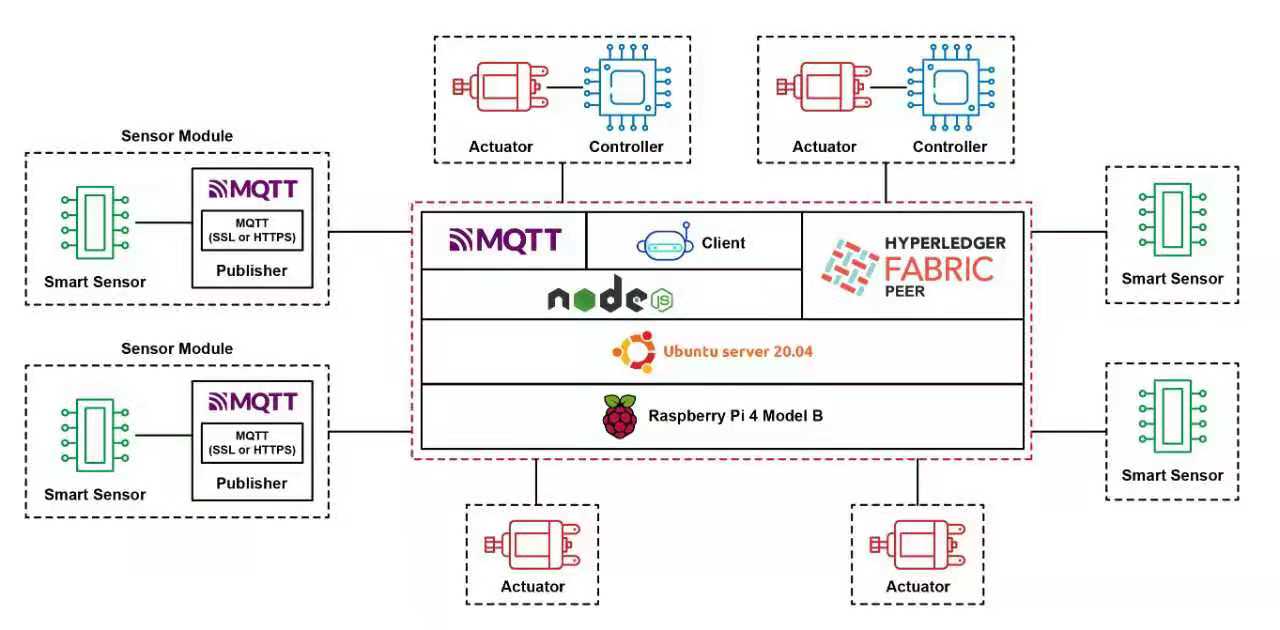}}
\caption{IoT Configuration Package over Raspberry Pi}
\label{fig:iot-config-r}
\end{figure*}

Our model cofiguration is shown in Figure \ref{fig:iot-config-r}. The system may be split into three levels. Physical layers comprised the system board itself and the smart sensors linked directly or across the network through MQTT broker. Our IoT terminal used a Raspberry Pi 4 Model B (8GB RAM). The IoT terminal runs on 64-bit ARM. The software layer includes the Ubuntu 20.04 stable version with long-term support (LTS). The next layer is the services layer where we have the Fabric and Nodejs along side the MQTT, Fabric Software Development Kit and custom firmware to interact with sensors. No Hyperledger fabric binaries and Docker Images are supported by default for arm-based systems. For our IoT terminal, we got the source code and compiled it [Figure \ref{fig:pi-docker}]

Our model decrease inefficiencies, increase security, and improve transparency for all parties involved, while also allowing safe machine-to-machine transactions to take place. The Hyperledger Fabric Policies make us capable of deriving a complete decentralized access control over the device itself. 

\section{Prototype Network}

\begin{figure*}[t]
\centerline{\includegraphics[scale=0.35]{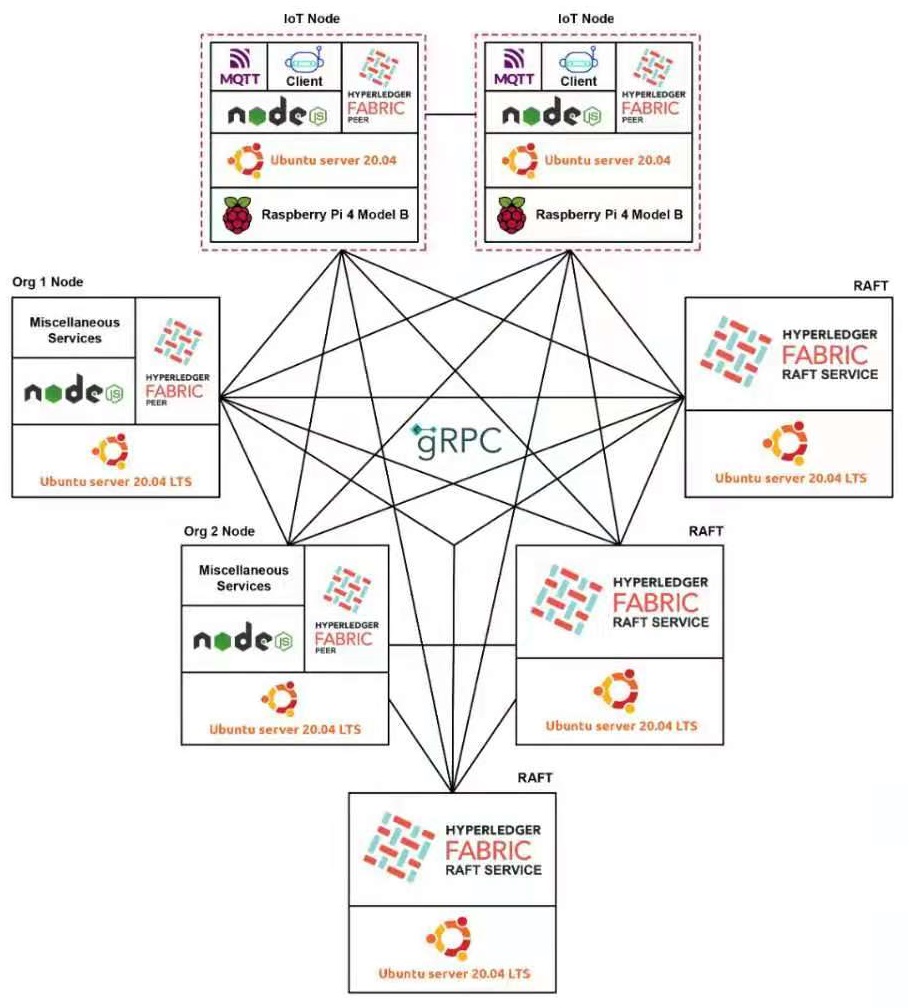}}
\caption{Hyperledger Fabric Blockchain Network for IoT}
\label{fig:blockchain-network}
\end{figure*}

The experimental network for IoT-blockchain integration includes four organizations: Org1, Org2, IoT, and Orderer [Figure \ref{fig:blockchain-network}. IoT equipment, such as temperature sensors, and NH3 gas sensors are constantly monitoring the temperature, humidity and NH3 presence in a grain silo and transmitting the information to a smart contract that is now in operation. To secure the network communication, the each organization's TLS-CA server provides TLS (Transport Layer Security) to the CA (Certification Authority) server, peers and all the client connections to the network. The organization's CA server provides X509 certificates to all components, users and clients  in its blockchain network. It runs on Xen Hypervisor with Xeon E5-2678-V3 X2 and 128 GB RAM. Each machine has two cores and two gigabytes of RAM and Ubuntu server 20.04 LTS as base OS. We used the Fabric 2.2.3 LTS as the blockchain network. Table \ref{tab:hw-specifications} lists the network nodes of our experimental network.

\begin{table}[h]
\caption{\label{tab:hw-specifications} Experimental Network Machines}
\begin{tabular}{ll}
\hline
\textbf{Host}  & \textbf{IP Address} \\ \hline
IoT-TLS-CA     & 10.0.5.10  \\ \hline
IoT-CA         & 10.0.5.20  \\ \hline
peer@IoT       & 192.168.10.60 \\ \hline
Org1-TLS-CA    & 10.0.3.10  \\ \hline
Org1-CA        & 10.0.3.20  \\ \hline
peer@org1      & 10.0.3.30   \\ \hline
Org2-TLS-CA    & 10.0.4.10  \\ \hline
Org2-CA        & 10.0.4.20  \\ \hline
peer@org2      & 10.0.3.30   \\ \hline
Orderer-TLS-CA & 10.0.100.10 \\ \hline
Orderer-CA     & 10.0.100.20 \\ \hline
solo@orderer   & 10.0.100.50 \\ \hline
\end{tabular}
\end{table}

\section{Working Process}

The suggested architecture is intended to provide for safe administration of activities in the intelligent technological environment. These activities comprise system startup, sensors and actuator data collection, and reliable customer control over the device to perform desired operations. All these procedures are explained briefly in the following sections.

\subsection{Device Registration}

All players in the Hyperledger Fabric blockchain network must get network membership in the form of an x509 certificate. Our network's IoT organization has its own specialized Membership Service Provider entity. As shown in the table, the entity maintains two servers, IoT-TLS-CA and IoT-CA. The IoT-TLS-CA server produces TLS certificates for our terminal. The IoT terminal obtained its TLS certificate and membership certificate as a peer from the IoT-TLS-CA server. These certificates are required to join the network. The required certificate needs to provide to add in the configuration block of the blockchain netowrk. The device now can join the blockchain netowork by joining the required channel.

\subsection{Device Access Control}
Fabric Policies allow us to restrict access to our resources to those required by the organization. Our Internet of Things (IoT) devices, for example, are sensitive. The consortium decided that the chaincode may be invoked only by the administrator of the IoT organization, and the chaincode will only endorse on the IoT Organization terminals etc. The Hyperledger Fabric policies provided the basic decentralized control over the device decided and agreed by the consortium.

\subsection{Chaincode Access Control}
Hyperledger Fabric provided an explicit library to apply access control over the cahincode. It needs that the access control must be explicitly implemented in the chaincode itself. We may encode the access control into the chaincode using the Hyperledger Fabric client identification library. We must manually import this library into our chaincode since it is not included in the Hyperledger Fabric blockchain network's basic functionality.

\subsection{Deploy Chaincode}
We developed and deployed a Silo Monitoring chaincode on the IoT terminal as well as other peer nodes in the network. The chaincode may be found on GitHub, where it can be downloaded [https://github.com/adnanjee/silomonitor.git]. ARM architecture is used in the IoT terminal, therefore we must package the ARM architecture's chaincode on the IoT device itself in order for it to function properly.

\section{Results and Discussion}

In Internet of Things devices such as the Raspberry Pi, the total performance of the device is also dependent on the memory type we use, such as the class of memory card, USB storage, or SSD storage in USB ports, and the amount of memory we use. We made a basic performance benchmark on our device to show its performance on the netowrk.

\subsection{Performance Benchmark}

\begin{figure*}[t]
\centerline{\includegraphics[scale=0.25]{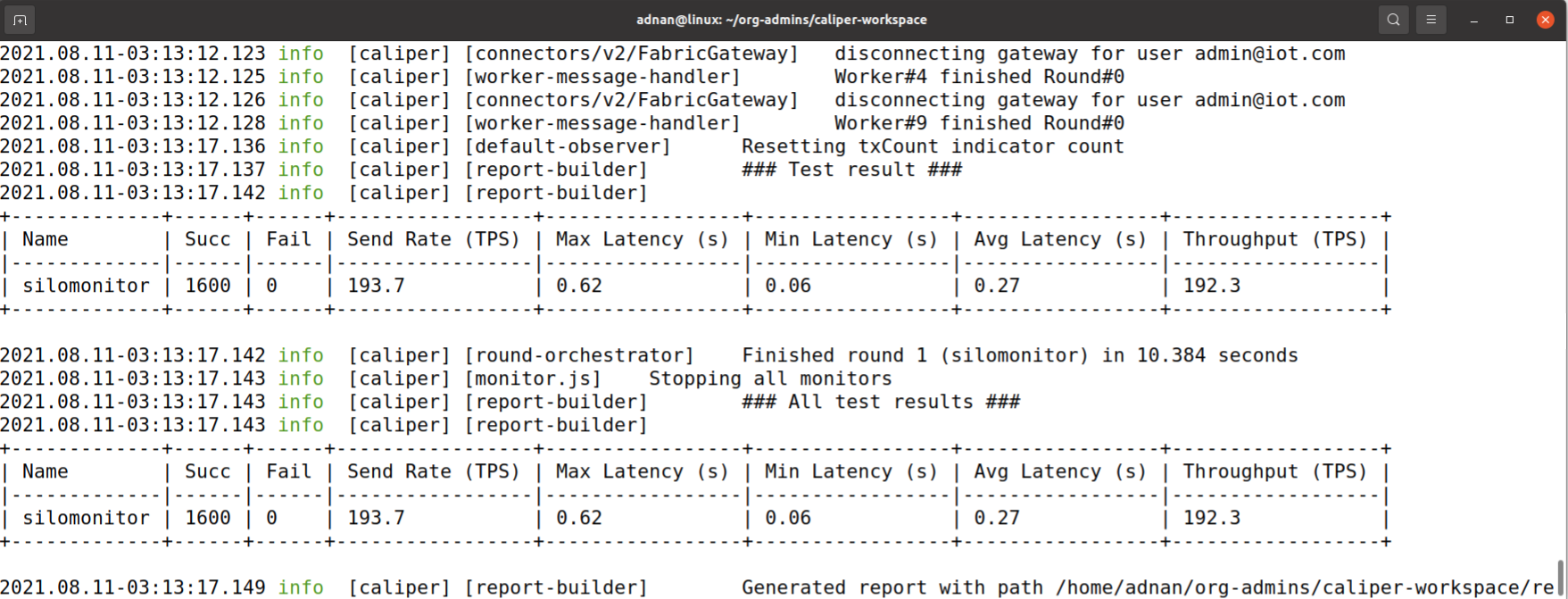}}
\caption{Performance Benchmark on IoT Terminal}
\label{fig:pi-performance}
\end{figure*}

Several benchmarking tools available, such as Hyperledger Caliper, may be used to determine performance. At this moment, we are just monitoring the most basic performance indicators using the Hyperledger Caliper tool, which is available for free on the Internet. In terms of the reading temperature function, we are getting about 192 transactions per second. The max latency was 0.62 seconds during extensive testing, the minimum latency was 0.06 seconds, and the average latency was 0.27 seconds [\ref{fig:pi-performance}].

\subsection{Security Factor}
Security considerations are concerned with the critical pillars that influence the adoption of blockchain in the Internet of Things architecture. The implementation of appropriate access control and authentication frameworks allows businesses to identify Internet of Things devices, isolate compromised nodes, guarantee the integrity of data, authenticate users, and permit various degrees of data access, among other things. Here the access control system comes handy. As our IoT device is a direct node of the blockchain network, we have a completely decentralized control over our device.

The access control system is derived from consortium planning. To put it simply, the consortium determines the Hyperledger Fabric regulations and rules that govern the whole system. The Hyperledger Fabric regulations describe how to access and update network and channel parameters. The consortium sets the first rules, since it is in charge of providing fine-grained access control throughout the blockchain network. Fabric Policies allow us to restrict resource access to meet organizational needs. For example, our IoT devices are sensitive. The consortium decided that only the IoT organization's administrator may use the chaincode. Users can set policies for chaincode execution in Hyperledger Fabric. These endorsement policies describe which peers must approve a transaction before it can be recorded.

\subsection{Data Privacy and Data Trading}

It is significant because it controls how data is used and who has access to it. When it comes to data privacy, blockchain technologies have taken the lead. The Hyperledger Fabric provides the library to explicitly restrict access to the chiancode. Hyperledger Fabric client identification library allows us to implement access control into chaincode. This library is not part of the core Hyperledger Fabric blockchain network functionality and must be manually imported.

This access control over the chaincode provides and additional layer of privacy and improve the data trading model with partner organizations in combination with the Blockchain Channel. We may specifically allows a certain user from a certain organization to control the chaincode as they are using the grain silo for that moment.

\subsection{Data integrity}
Data assurance, completeness, consistency, and dependability are the primary goals of this process, which is carried out across the whole data life cycle. The cloud services that are connected to IoT devices have caused the majority of the problems with data integrity that have been observed. The blockchain's fundamental architecture is a sequence of increasing blocks hashed using cryptography to ensure data integrity while allowing a completely decentralized system. Aside from that, out IoT terminal function as a direct blockchain node, processing and storing data without the need for human involvement or transmission to another system which makes the data more reliable.

\subsection{Data Governance}
A data strategy, rules, standards, procedures, and matrices are all defined as part of the data governance process. It is engaged in the monitoring of the completion of data management projects and providing data management services. In order to manage privacy and other standards for each area of business, data regulations are constantly being revised and updated. 

\section{Conclusions \& Future Work}

The Internet of Things in the industrial sector modernizes smart industries by integrating the latest technologies. Throughout this article, we propose a blockchain-based architecture for the Internet of Things devices network. It was shown how to configure the Raspberry Pi 4 so that it may connect to the blockchain network and act as a blockchain node. We look at the standard Internet of Things architecture and evaluate the security and privacy issues that come with IoT deployments. We later discussed that how our proposed configuration eliminates many of those issues and facilitates the organizations to control the IoT devices with data security and privacy. Future work should include an extensive network of IoT devices such as Raspberry Pi, Odroid XU4, Asus Tinker Board, and Nvidia Jetson nano. The system must assess hardware capabilities, processor and memory limits, power consumption, network consensus methods, scalability, and network latency. So we may select where to install full-scale servers or edge servers, or leverage our existing IoT devices for different purposes, including consensus networks.

\section{Data Availability}

This article does not fall under the category of data sharing.

\section{Conflicts of Interest}

The authors are not affiliated with any entity that has a financial stake in the subject matter addressed in the article, either directly or indirectly.

\section{Funding Statement}

The authors want to express their gratitude for the assistance given by China's National Key R\&D Program (No. 2018YFC1604000).

\section*{References}

\bibliography{manuscript}

\end{document}